# A Framework for Leveraging Human Computation Gaming to Enhance Knowledge Graphs for Accuracy Critical Generative AI Applications

Steph Buongiorno
*Guildhall*
*Southern Methodist University*
Dallas, United States
sbuongiorno@smu.edu

Corey Clark
*Computer Science and Guildhall*
*Southern Methodist University*
Dallas, United States
coreyc@smu.edu

*Abstract*—External knowledge graphs (KGs) can be used to augment large language models (LLMs), while simultaneously providing an explainable knowledge base of facts that can be inspected by a human. This approach may be particularly valuable in domains where explainability is critical, like human trafficking data analysis. However, creating KGs can pose challenges. KGs parsed from documents may comprise explicit connections (those directly stated by a document) but miss implicit connections (those obvious to a human although not directly stated). To address these challenges, this preliminary research introduces the GAME-KG framework, standing for "Gaming for Augmenting Metadata and Enhancing Knowledge Graphs." GAME-KG is a federated approach to modifying explicit as well as implicit connections in KGs by using crowdsourced feedback collected through video games. GAME-KG is shown through two demon- strations: a Unity test scenario from Dark Shadows, a video game that collects feedback on KGs parsed from US Department of Justice (DOJ) Press Releases on human trafficking, and a following experiment where OpenAI's GPT-4 is prompted to answer questions based on a modified and unmodified KG. Initial results suggest that GAME-KG can be an effective framework for enhancing KGs, while simultaneously providing an explainable set of structured facts verified by humans.

## I. INTRODUCTION

Large language models (LLMs) have demonstrated remarkable capabilities in performing various natural language processing (NLP) tasks, including extracting semantic information and inferring cohesive text responses based on human input [10]. However, their application within accuracy critical domains–such as human trafficking data analysis–is limited due to a lack of explainability in results. An LLM's mechanism for generating a response is opaque and may contain unwanted biases or "hallucinations" (generated text that is semantically possible but factually incorrect) [12]. In accuracy critical domains, it may be too high-stakes to trust an LLM generated response without the ability to validate the process for generating it. For example, LLM generated claims that a person broke a law could be detrimental if they contain hallucinations, and the process for arriving at this claim has not been validated. Explainability is crucial for trusting LLM generated results and validating them when needed [6], [7].

Providing LLMs with external, structured representations of facts in the form of knowledge graphs (KGs) has proven useful for addressing this limitation [8]. KGs are graphs of data that collect and convey knowledge. They represent information as structured relations between entities, which make a "domain model," along with schematic information, which make up an "ontology" [15]. Information can be retrieved from KGs to augment LLM generated responses in an explainable way [11], [18]. Human analysts can visually interpret the connections between nodes of the graph and view the structured path of facts from which information was derived [3].

While using KGs to augment LLM responses has proven beneficial, obtaining KGs that represent domain information poses challenges. KGs can be parsed from documents, but capturing the many entity relationship from within text may not be feasible with computational approaches alone [21]. For instance, a KG parsed from text may include explicit relations between entities (e.g. semantic relations as directly expressed in words), but may not include implicit relations (e.g. relations that may be clear to a human, but not to the computer). In other cases it may be desirable to modify the explicit relationships of a KG, given that the connections might be incorrect.

For these reasons, this research introduces the GAME-KG framework (standing for "Gaming for Augmenting Metadata and Enhancing Knowledge Graphs"). GAME-KG is a federated approach to modifying KGs that facilitates the collection of explicit as well as implicit knowledge. It leverages Human Computation Gaming (HCG)–a method of collecting feedback from crowds through video games–to modify and validate KGs [4]. GAME-KG guides the process of parsing a KG to presenting the data to the player and collecting feedback.

This research explores GAME-KG's potential across two demonstrations. The first demonstration presents a video game test scenario from a HCG, Dark Shadows, designed using

This work has been funded by National Institute of Justice (NIJ). Invaluable insight was provided by the anti-human trafficking agency, Deliverfund. Thanks to BALANCED Media | Technology for developing the Unity demo.

the GAME-KG framework.[1] Dark Shadows is a film noir-style mystery game that collects player feedback for modifying and validating KGs parsed from the US Department of Justice (DOJ) Press Releases on human trafficking. This demonstration presents a drag-and-drop mechanic, showcasing how a video game can be used to collect feedback and modify KGs. For the second demonstration, OpenAI's GPT- 4 is asked questions related to the human trafficking press releases and is prompted to provide answers based on the original KG parsed from text, and then a human modified KG in which connections were added that encode implicit relationships between entities. The initial results from these demonstrations suggest that a HCG designed via GAME- KG can be an effective way to collect human feedback, and that information retrieved from KGs can be used to augment LLM responses in an explainable way, which is important for accuracy critical domains like human trafficking data analysis.

## II. BACKGROUND

This section provides a brief background on KGs and explainability before describing the utility of crowdsourcing feedback for modifying and validating KGs.

### A. Knowledge Graphs as Explainable Domain Models

A KG is a graph composed of nodes and edges, where nodes represent entities, and the edges between nodes convey the semantic relationship between entities. KGs model domains in the sense that their relations encode information on a subject. For example, a KG representing the domain of data science might include relationships between scientific software packages, problem solving strategies, and metrics. Whereas, a KG for GIS might include relations between geographical entities, documents describing terrain, and coordinate systems. Structured graphs have the added benefit of explainability. LLMs are complex, "black-box" systems and responses are generated without insight into the model's inner workings. Their high complexity makes model interpretation challenging [20]. Yet, explainability is important because it can make evident unintended biases, or it can reveal the information on which a response was based. Presented with a KG, human analysts can visually interpret the connections between nodes and read the structured path of facts from which information was derived [3]. This approach gives insight into the individual entity relationships that encode knowledge, and it supports the development of more reliable and trustworthy models for critical applications.

### B. Gamification and Human Computation Gaming (HCG)

Obtaining KGs that represent both explicit and implicit knowledge is not a trivial task, and a KG parsed from a document using computational methods alone may benefit from human feedback to connect or validate the relationships between nodes. Crowdsourcing has been used to construct KGs manually, as in Wikidata [19]. However, motivating a crowd to provide feedback poses challenges. One effective method for engaging and motivating crowds is through gamification, a proven mechanism for engaging individual users and collecting large amounts of heterogeneous knowledge to achieve a common goal [14]. Gamification has been widely used across diverse subjects–including education, healthcare, and marketing–to drive participation and motivate players, which may result in a higher feedback yield [5], [22].

[1]A Dark Shadows Unity demo, code, prompts, knowledge graphs, and examples can be found on our GitHub repository.

HCG–a method that gamifies crowdsourcing using video games–leverages video game mechanics that are designed to engage players while simultaneously collecting the feedback desired for a given task. In addition, HCG can be a desirable approach because it provides the opportunity to tap into a large player base. Globally, there are over three billion "gamers," with a predicted annual +3.7% player growth [13].

### C. Crowdsourcing Implicit Knowledge

Crowdsourcing has been used to collect knowledge that may be obvious to a human but not a machine [1]. Such insight may be especially useful for processing documents, as text often conveys not only explicit information (information directly expressed in words) but also implicit information (information that can be inferred, but is not directly stated) [17].

Implicit information is derived between a claim and background knowledge, as exemplified by Table I. It is common for documents to omit information that is assumed to be evident to the recipient, such that only parts of a message may be explicitly stated in words [1]. In result, a KG parsed from a document may not capture the many implied entity relationships expressed in a document. While LLMs have demonstrated the ability to perform inference and emulate common-sense reasoning responses, this ability is not exhaustive as a LLM may not possess the domain knowledge needed for certain implied information, indicating a need for crowdsourcing human feedback [1].

**Claim:**
John Doe trafficked humans.
**Background Knowledge:**
It is illegal to traffick humans.
**Implicit Knowledge:**
John Doe broke the law.

TABLE I
A CLAIM, BACKGROUND KNOWLEDGE, AND THE RESULTING IMPLICIT KNOWLEDGE.

## III. THE GAME-KG FRAMEWORK

GAME-KG, shown by Figure 1, is a six-step federated approach to knowledge acquisition that leverages HCG to modify and validate KGs with the goal of enhancing them for critical, domain tasks and downstream generative AI applications. A HCG developed using the GAME-KG framework enables the collection of human feedback to modifying explicit as well as implicit entity relationships in a KG.

The six-steps of GAME-KG are as follows:

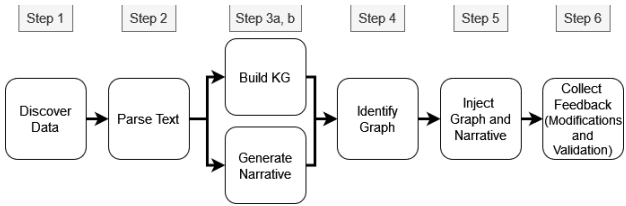

Fig. 1. The steps of the GAME-KG Framework.

- (Step 1: Discover Data) The document(s) that will be used to develop a domain-model are identified and collected.
- (Step 2: Parse Text) Entities and their relationships are parsed from the document(s).
- (Step 3a: Build KG) The parsed text is used to construct a KG. Entities are nodes and their relations are edges.
- (Step 3b: Generate Narrative) The parsed text is provided to an LLM instructed to generate a new, fictionalized narrative for the HCG. Research has shown that LLMs can generate stories for video games [2], [9]. For GAME-KG, this step is used to generate narrative that is more palatable for gameplay while using the original entity relationships. For human inspection, the content can be anonymized and fictionalized within the game narrative.
- (Step 4: Identify Graph) The entities and relations making up the KG can be presented to the player for inspection. If desired, these can be identified based on a scoring method, such as cosine similarity. This approach suggests certain subsections of the graph may benefit more from human feedback than others. If the cosine similarity between nodes is low, this could indicate that the connec- tions are possibly wrong. If the cosine similarity between nodes is high but the edges are not connected, this could indicate that a connection should be made. Humans can judge whether a connection between entities should be modified. This method for identifying entities and nodes can change based on need.
- (Step 5: Inject Graph and Narrative) The entities and relations as well as the fictional narrative are injected into the game.
- (Step 6: Collect Player Feedback (Modifications and Validation)) Each time a player suggests changes–based on explicit or implicit knowledge–these changes are saved as a weighted value. Connections between nodes with a low weight can be filtered out. This represents a federated approach, where based on consensus, the changes are retained or omitted.

IV. DEMONSTRATION I: GAME-KG TEST SCENARIO

This section presents a test scenario from the HCG, Dark Shadows. Dark Shadows is a film noir-style mystery game that leverages GAME-KG to collect player feedback that modifies and validates KGs parsed from US DOJ Press Releases on human trafficking. In preparation of presenting narrative to the player, a subsection of the KGs are provided to GPT-4 with instruction to generate a fictional and anonymized narrative that retains the original entity relations. The resulting narrative contains key information meant for human inspection, but fictionalizes it for player engagement. As an example, the generated narrative might feature fictional events, but contain real relationships between entities.

In Dark Shadows, the generated narrative is presented to the player as a fictional case briefing, as shown by Figure 2. Using a drag-and-drop mechanic resembling an interconnected web, or "evidence board", players can select entities from the case file and specify connections between them by dropping them into the web. Weight can be added to the human-made connections as desired for validation, supporting the ability to collect graphs and also filter out responses with a low weight. As a HCG, Dark Shadows's design is based on the premise that gamification–in this case, progressing through a fictional narrative–can motivate players into providing feedback [22]. To foster this engagement, Dark Shadows also uses a gen- erative, text-to-image model, Stable Diffusion, to generate in- game images matching the entities selected by the player [16].

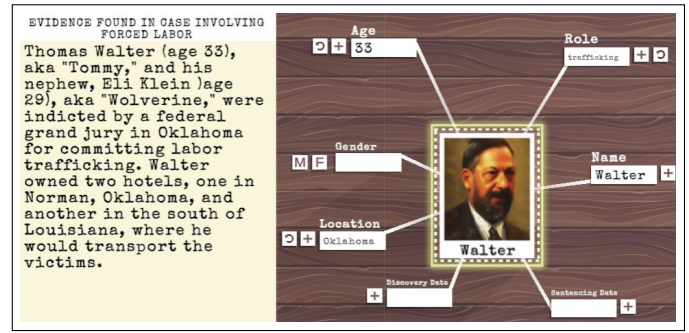

Fig. 2. Players are presented a fictional narrative that retains entity relations and can modify a KG using a web-like mechanic. The image of a "suspect" is generated with Stable Diffusion.

V. DEMONSTRATION II: PRELIMINARY RESULTS EVALUATION THROUGH Q&A

The web-like mechanic can be used to modify explicit or implicit entity relations in a KG. This preliminary research focuses on modifying implicit entity relations. It then compares GPT-4's responses as augmented by information retrieved from a human modified graph and an unmodified KG.

Table II shows an example human modification that is based on implicit knowledge.[2] In this example, Kizer, an identified trafficker, is explicitly described as breaking the Mann Act for transporting a trafficked victim across state boarders. Villaman is an accomplice to Kizer, but his relation to the Mann Act is not explicitly stated. To enhance the KG, a human connected Villaman and the Mann Act, as shown by Figure 3.

The modified KG was provided to GPT-4 for question-answering (Q&A). As shown by Table III, information retrieved from the modified KG can augment the generated response. This approach is beneficial for accuracy critical domains because structured representations of facts, like KGs, are explainable and can contain human validation.

---

[2]For additional examples, see our GitLab.

| |
|---|
| **Claim:** Kizer transported victims across state boarders. Villaman was an accomplice to Kizer. |
| **Background Knowledge:** The press release states Kizer broke the Mann Act when he transported a victim across state boarders. |
| **Implicit Knowledge:** Villaman also violated the Mann Act. |
| **Modified Connection:** Villaman, violated, Mann Act |

TABLE II
HUMAN MODIFIED CONNECTIONS BASED ON CLAIMS AND BACKGROUND KNOWLEDGE.

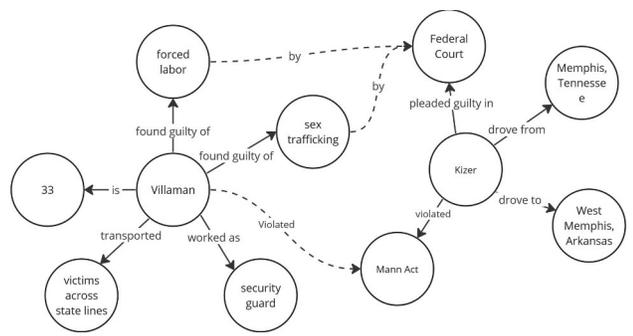

Fig. 3. An abstracted KG based on press release data. Solid lines represent explicit relationships, and dashed lines represent human modified connections.

This approach can also guard against hallucinations or unexplainable responses. When prompted to answer the same question based on the unmodified graph–in which Villaman is not connected to Mann Act–the LLM was instructed to inform the user that the KG does not contain the knowledge to answer the question, as shown by Table III

| |
|---|
| Question: What act did Villaman break? |
| Answer (modified KG): The Mann Act |
| Answer (original KG): The knowledge to generate an answer is not found. |

TABLE III
Q&A USING OPENAI'S GPT-4.

## VI. DISCUSSION AND FUTURE WORK

The GAME-KG framework serves as one step towards leveraging Human Computation Gaming (HCG) to modify KGs for accuracy critical generative AI applications, such as human trafficking data analysis. Obtaining KGs that have undergone human modification and validation may be especially desirable for such applications, as there is need for explainablility. In the future, research will need to be conducted to explore the additional benefits and risks of leveraging human feedback for modifying KGs to better understand situations where humans may contribute desirable biases, as well as undesirable biases, thus creating a greater foundational understanding of HCG leveraged for KG modification.